# HEAVY ELEMENTS IN THE LYMAN-$\alpha$ FOREST: ABUNDANCES AND CLUSTERING AT Z=3

D.S. WOMBLE, W.L.W. SARGENT, R.S. LYONS
*California Institute of Technology*
*Astronomy 105-24, Pasadena, CA 91125 USA*

**Abstract.** For the purpose of studying the properties of heavy-elements associated with the Lyman-$\alpha$ forest, we observed the gravitational lens Q1422+2309. We used the HIRES instrument on the W.M. Keck telescope to obtain a high-resolution, very high signal-to-noise spectrum of this $z = 3.63$ quasar; the spectrum covers wavelengths from below Lyman-$\beta$ up to the C IV emission line. Consistent with previous estimates, we find that a moderate fraction of the Ly-$\alpha$ forest clouds have been enriched with heavy elements at a level significantly below solar abundance. However, unlike the fairly uniform distribution of Ly-$\alpha$ forest lines, we show that the C IV absorption lines are clustered on large velocity scales.

## 1. Background

The spectra of high-redshift quasars show remarkable complexity in H I absorption at wavelengths shortward of Ly-$\alpha$ emission. The nature and composition of these ubiquitous Ly-$\alpha$ "forest" lines remains as a prominent topic in cosmology. For more than a decade, the consensus has prevailed that high redshift Ly-$\alpha$ forest lines are due to an intergalactic population (*cf.* Sargent et al. 1980). Motivated to ask whether these Ly-$\alpha$ forest clouds have primordial or enriched heavy element abundances, we report on a very sensitive study of C IV absorption corresponding to the Ly-$\alpha$ forest in Q1422+2309.

Several measurements and upper limits on the metallicity of the Ly-$\alpha$ forest have been reported in the literature (Norris et al. 1983, Meyer & York 1987, Lu 1991, Tytler & Fan 1994). Using a variety of techniques, these results have provided some evidence for heavy element enrichment in the forest. However, it is not clear what fraction (if any) of the population show primordial abundances and if the enriched clouds originate from a physically distinct class of object (such as extended galaxy halos). The exceptional high-resolution facilities on the Keck 10-m telescope provide a timely opportunity to explore the properties of metals associated with the Ly-$\alpha$ forest. We have obtained a spectrum, of unprecedented quality, on Q1422+2309 for this purpose.

The quasar 1422+2309 is a bright, high-redshift (V=16.5, $z_e = 3.63$) gravitational lens (Patnaik et al. 1992). This source is composed of four unresolved components located within an angular diameter of 1.3 arcseconds. Hammer et al. (1995) report a redshift for the lens of $z = 0.647$ however we have not confirmed this measurement – owing to the absence of any associated intervening absorption lines in the quasar spectrum. The density of the Ly-$\alpha$ forest in this high-$z$ quasar makes it difficult to identify any $z < 1$ metal line systems which might signify the redshift for the lensing mass.

## 2. Observations

We observed Q1422+2309 using the HIRES spectrograph on the Keck 10-m telescope during June 1994 and Spring 1995. The data were taken through a 0.86 by 7 to 14 arcsec slit resulting in a resolution of 6.6 km s$^{-1}$ FWHM. Because the position angle of the slit rotates during an exposure (since Keck is Alt-Az), the spectrum contains significant amounts of light from the three brightest images of this gravitational lens.

We have achieved a typical signal to noise ratio of 150:1 per resolution element from a total exposure time of 25000 seconds. By using several different instrumental setups, the spectrum has complete wavelength coverage over $\lambda\lambda 3650$–7150 (over 94000 pixels) with no inter-order gaps. At the redshift of the quasar, this spectrum extends from below 912Å to just short of the broad C IV emission line peak.

Excluding a few isolated regions, we reach a limiting sensitivity under $10^{12}$ cm$^{-2}$ in the column densities of both C IV (unresolved) and Ly-$\alpha$ (for $b \sim 20$ km s$^{-1}$) over most of the spectrum.

## 3. Results

In principle, the extended (lensed) nature of the background source may adversely affect intervening absorption line statistics because of cloud covering factors and inhomogeneous density distributions. The maximum angular separation of lensed images, at a lens redshift of 0.647, corresponds to an impact parameter of 5 $h^{-1}$ kpc ; we are using H$_o = 100h$ km s$^{-1}$ Mpc$^{-1}$, q$_o$=0.5 herein. Fortunately, this length is much smaller than recent esti-mates for the spatial extent of Ly-$\alpha$ clouds (Dinshaw et al. 1995) and the inferred diameters of galaxies which produce typical C IV absorption (Sargent et al. 1988). If, for example, a Ly-$\alpha$ forest cloud did not completely cover the background source, we would expect to see a highly non-Voigt line profile due to a zero-point offset in the residual flux. No such line profiles have been observed. The question of density fluctuations relative to the individual images has not yet been addressed and it may remain as an insoluble problem.

Concentrating on the portion of the spectrum between Ly-$\beta$ and Ly-$\alpha$ emission, $2.95 \leq z_a \leq 3.61$, we find that a significant fraction of the strong Ly-$\alpha$ forest lines have detectable C IV absorption. Using an optical depth criterion to select H I lines with $\tau \geq 5$, we count 66 Ly-$\alpha$ lines over this redshift path. For a Doppler parameter, $b = 30\,\mathrm{km\,s^{-1}}$, this limit corresponds to a column density, N(H I)$\geq 2 \times 10^{14}\,\mathrm{cm^{-2}}$. At the corresponding positions of C IV absorption, we detect 26 metal-line systems; for these purposes, all C IV components spread over $\Delta v \leq 500\,\mathrm{km\,s^{-1}}$ are counted as a single redshift system. We also detect an additional 4 C IV systems at redshifts corresponding to Ly-$\alpha$ lines which have apparent optical depths below our selection threshold. These latter systems, and many components of C IV absorption appear to have different velocity distributions than their associated Ly-$\alpha$ lines. Allowing that the H I and C IV ions may occupy physically distinct regions in the absorbing clouds, we find that 40–45% of the Ly-$\alpha$ forest lines with $\log$ N(H I)$\geq$14.30 have been enriched with heavy elements. This fraction is roughly consistent with the values of 60% (Tytler et al. 1995) and 50% (Cowie et al. 1995) for logN(H I)$\geq$14.5; note that we see 30/66 C IV systems compared with 13/22 and 15/31, respectively.

For the majority of these H I systems, the Ly-$\alpha$ lines are too saturated to obtain reliable column densities. In the six weakest systems, we obtained accurate measures of the C IV and H I column densities using Voigt profile fitting of the lines. In these systems, the median difference between the redshifts of the C IV and Ly-$\alpha$ components corresponds to a velocity offset of $35\,\mathrm{km\,s^{-1}}$. The mean relative abundance, $\langle \log \mathrm{N(C\,IV)}\,/\,\mathrm{N(H\,I)} \rangle = -2.65 \pm 0.22$.

In order to convert to an absolute abundance, we must make an *a priori* assumption on the ionization state of the gas. Adopting C/C IV $=10$ and H/H I $=10^4$, we obtain a typical abundance of [C/H]$=-2.3$ at $\langle z \rangle \simeq 3.1$; this value is highly dependent on the (unknown) ionizing conditions in the clouds. For comparison, Lu (1991) found [C/H]$\sim -3.2$ for relatively strong Ly-$\alpha$ lines whereas Tytler & Fan (1994) placed a limit of [C/H]$\leq -2.0$ at lower H I column densities and Tytler et al. (1995) claim [C/H]$> -2.5$ for logN(H I)$\geq$14.5. Given uncertainties in the ionization corrections, our value is entirely consistent with previous abundance estimates.

In Figure 1, the lower panel shows the portion of normalized quasar spectrum for Ly-$\alpha$ lines with $2.95 \leq z_a \leq 3.62$. On the same scale, the upper panel shows the distribution of C IV column densities as a function of redshift; these values were obtained by least-squares Voigt profile fitting with the minimum number of velocity components needed to get a satisfactory fit. It is visually apparent from this Figure that the C IV systems are clustered. The tight groups of lines are clustered on a velocity scale less than $500 \, \mathrm{km \, s^{-1}}$; this scale has already been well established by lower-resolution studies. It is also evident by eye, that the aggregate set is lumped on scales approximating several thousand $\mathrm{km \, s^{-1}}$.

For this database of 146 C IV redshifts, we have evaluated the two-point correlation function in comoving coordinates. Figure 2 (top) shows this correlation function for velocities less than $1000 \, \mathrm{km \, s^{-1}}$, whereas the bottom panel extends out to redshift-pair separations of $10000 \, \mathrm{km \, s^{-1}}$. It can be seen that there is significant structure both below $500 \, \mathrm{km \, s^{-1}}$ and in the vicinity of $4500 \, \mathrm{km \, s^{-1}}$ (at the $5\sigma$ level). These structures correspond to comoving distances of $\leq 2.5 h^{-1}$ Mpc and $25 \, h^{-1}$ Mpc, respectively. The structure seen on smaller velocities ($\Delta v \leq 500 \, \mathrm{km \, s^{-1}}$) is likely to stem from galaxy-galaxy correlations although some component is also certainly due to the motion of clouds within individual galaxy halos. On the much larger scale, Sargent et al. 1988 saw some evidence for this structure in the distributions of redshifts in their lower-resolution C IV survey. The higher sensitivity of the present survey is probably responsible for the increased significance of such characteristics. It is interesting to note that the 25 $h^{-1}$ Mpc scale length at $z \simeq 3$ is comparable to the local cluster-cluster correlation length.

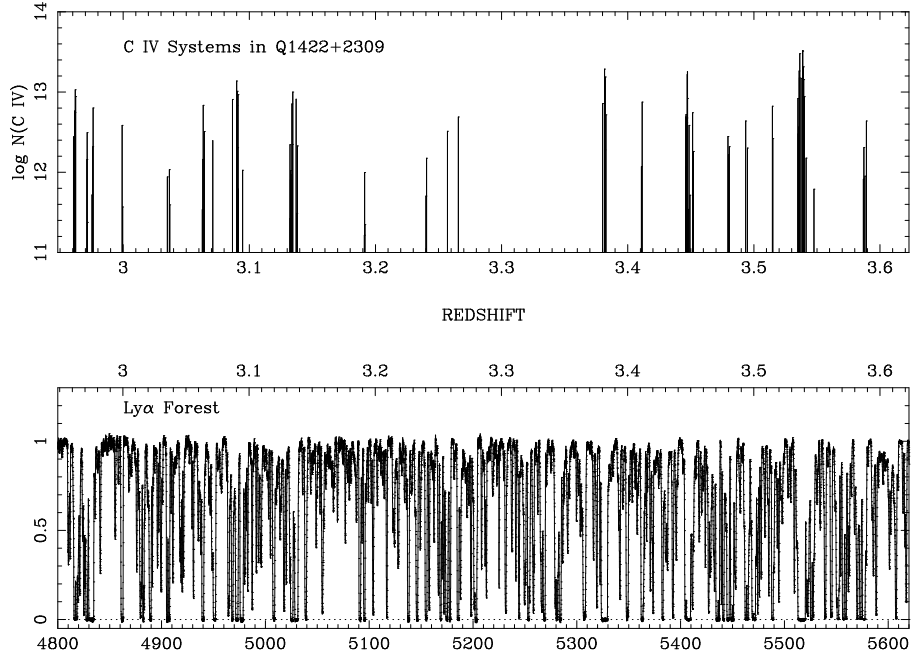

Figure 1. Lower panel shows normalized spectrum in Ly-α forest with $2.95 \leq z_a \leq 3.62$; upper panel shows the distribution of C IV column densities as a function of redshift (same x-scale). Note the apparent clustering of C IV systems both on small and large-scales.

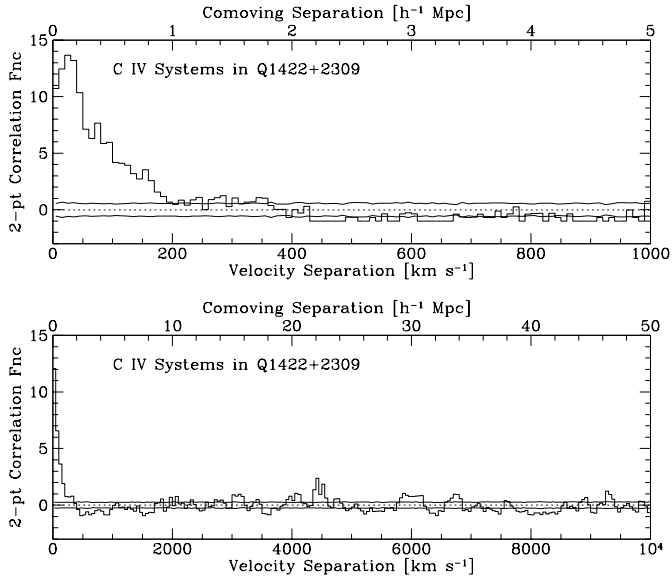

Figure 2. Two-point correlation functions for 146 C IV redshifts shown in Fig 1; pair separations denoted in both velocity and comoving distance for small and large-scales.